\documentclass[twocolumn,english,aps,reprint, superscriptaddress,showpacs,longbibliography,showkeys, nofootinbib]{revtex4-2}
\usepackage{amsmath,amssymb,bbm,mathrsfs,bm,braket,color,graphicx,comment,xcolor,dsfont,multirow}
\usepackage[colorlinks,citecolor=blue,urlcolor=blue,linkcolor = blue]{hyperref}
\usepackage[mathscr]{euscript}

\usepackage{subfigure}
\usepackage{multirow}

\begin{document}
\title{Applications of Universal Parity Quantum Computation}
\author{Michael Fellner}
\affiliation{Institute for Theoretical Physics, University of Innsbruck, A-6020 Innsbruck, Austria}
\affiliation{Parity Quantum Computing GmbH, A-6020 Innsbruck, Austria}
\author{Anette Messinger}
\affiliation{Parity Quantum Computing GmbH, A-6020 Innsbruck, Austria}
\author{Kilian Ender}
\affiliation{Institute for Theoretical Physics, University of Innsbruck, A-6020 Innsbruck, Austria}
\affiliation{Parity Quantum Computing GmbH, A-6020 Innsbruck, Austria}
\author{Wolfgang Lechner}
\affiliation{Institute for Theoretical Physics, University of Innsbruck, A-6020 Innsbruck, Austria}
\affiliation{Parity Quantum Computing GmbH, A-6020 Innsbruck, Austria}
\date{\today}

\begin{abstract}
We demonstrate the applicability of a universal gate set in the parity encoding, which is a dual to the standard gate model, by exploring several quantum gate algorithms such as the quantum Fourier transform and quantum addition. Embedding these algorithms in the parity encoding reduces the circuit depth compared to conventional gate-based implementations while keeping the multi-qubit gate counts comparable. We further propose simple implementations of multi-qubit gates in tailored encodings and an efficient strategy to prepare graph states.
\end{abstract}

\maketitle

\section{Introduction}
In recent decades, there has been an enormous effort to develop novel strategies for quantum computation \cite{Deutsch1992, Shor1997, Bernstein1997, Grover1996, Reck1994, Barenco1995_Elementary, Mosca1999}, including measurement-based \cite{Raussendorf2001, Raussendorf2003} or adiabatic \cite{Albash2018} quantum computation, complementing the gate based paradigm of quantum computation. In Ref.~\cite{universality_paper}, we proposed universal quantum computation in the Lechner-Hauke-Zoller (LHZ) scheme \cite{Lechner2015} as a dual to the conventional gate model. Recent achievements in quantum hardware development on various qubit platforms \cite{henriet2020quantum, saffman2010quantum, bloch2008many, bernien2017probing, cirac1995quantum, blatt2012quantum, kielpinski2002architecture, jaksch2000fast} might soon allow for experimental realizations of well-known quantum algorithms for reasonable system sizes. Nevertheless, a fundamental challenge of state-of-the-art quantum devices remains the interqubit connectivity on quantum chips \cite{Hazra2021EnhancedConnectivity, Tadokoro2021QuantumDotAddressability}. This is especially pressing because a long-range and dense (ideally all-to-all) connectivity is a crucial ingredient for many key quantum algorithms, unless algorithm-specific preprocessing steps are performed \cite{Holmes2020_impact}. 
Unless efficient native implementations for long-range interactions as in, for example, the surface code \cite{Fowler2012, Beverland2022} exist, the connectivity problem is often dealt with by utilizing resource-intensive SWAP gates which, apart from requiring error-prone two-qubit gates, render a parallelization of gates difficult. Although there exist several quantum routing techniques \cite{Bapat2022, Devulapalli2022}, this is in particular problematic for scalability of devices beyond the noisy intermediate scale quantum (NISQ) era \cite{Preskill_2018}. 

An alternative way to address the connectivity issue and, as a side effect, allow for the native implementation of higher-order interactions, was introduced with the parity encoding \cite{Lechner2015, ender2021compiler}. The parity encoding maps $n$ logical qubits (with operators $\tilde\sigma$) to ${K>n}$ physical qubits (parity qubits, with operators $\sigma$), encoding the relative alignment (parity) along the $z$-axis of two or more logical qubits, such that for any state $\ket{\psi}$ in the code space
\begin{equation}\label{eq:parity}
    \tilde\sigma_{z}^{(i)}\tilde\sigma_{z}^{(j)}\ket{\psi} = \sigma_z^{(ij)}\ket{\psi}.
\end{equation}
Here, the superscripts correspond to qubit labels. In order to deal with the additional degrees of freedom in the physical Hilbert space, parity constraints of the form
 \begin{equation}\label{eq:constraints}
     C_{l}\ket{\psi}:=\sigma_{z}^{(l_{1})}\sigma_{z}^{(l_{2})}\sigma_{z}^{(l_{3})}\,[\sigma_{z}^{(l_{4})}]\ket{\psi}=\ket{\psi}
 \end{equation}
are introduced as stabilizers of the code space, which is also referred to as the constraint-fulfilling subspace $\mathcal{H}_\text{CF}$. The indices $l_i$ represent pairs of logical qubits. In every constraint, each logical index must occur zero or an even number of times. The square brackets around $\sigma_z^{(l_4)}$ indicate that a constraint can contain either three or four qubits. The special case of the parity encoding involving all $n(n-1)/2$ two-body terms (parity qubits) for $n$ logical qubits is known as the LHZ architecture \cite{Lechner2015}.  It is possible to extend this by including physical qubits representing single logical qubits such that
\begin{equation}
    \tilde\sigma_{z}^{(i)} =  \sigma_z^{(i)}.
\end{equation}
We refer to these qubits as \textit{data qubits} in the following. The presence of data qubits in the parity encoding ensures that the physical qubits uniquely define the state of the logical qubits\footnote{Note that two-body parity qubits alone determine the logical state only up to a global spin flip.}.
A variant of the LHZ architecture involving all data qubits has recently been shown to provide a universal gate set \cite{universality_paper}, based on the logical operators
\begin{align}
  \tilde R_x^{(i)}(\alpha) &=\exp\left(-i\frac{\alpha}{2} {\sigma}_{x}^{(i)}\textstyle\prod_{j<i}\sigma_x^{(ji)}\prod_{j>i}\sigma_x^{(ij)}\right)\label{eq:logical_rx}\\
  \tilde R_z^{(i)}(\alpha)&=\exp\left(-i \frac{\alpha}{2} \sigma_z^{(i)}\right)= R_z^{(i)}(\alpha)\label{eq:logical_rz}\\
      \text{C}\tilde{\text{P}}^{(i, j)}_{\phi} &= R_z^{(i)}\left(\frac{\phi}{2}\right)R_z^{(ij)}\left(-\frac{\phi}{2}\right)R_z^{(j)}\left(\frac{\phi}{2}\right) \label{eq:logical_cphase}.
\end{align}
Logical operators (with a tilde) act on the logical qubits defined in the constraint-fulfilling subspace and commute with all constraint operators. In contrast to that, physical operators (without a tilde) do not necessarily preserve the constraint-fulfilling subspace. 
The $\tilde R_x$ operators require chains of CNOT gates due to the product of Pauli operators in the exponent (see for example  Refs.~\cite{Cowtan2020, NielsenChuang2011, Steudtner2019} for further background on exponentiating products of Pauli matrices), while the other logical operators can be implemented with physical single-qubit operations only. The set of physical qubits that are involved in $\tilde R_x^{(i)}$ (i.e.\ the set of all physical qubits containing the logical index $i$) is referred to as the \textit{logical line} associated with qubit $(i)$.

In this work, we study the implementation of several essential quantum algorithms in this scheme and find that, depending on the algorithm, the parity scheme can show an advantage in circuit depth or multiqubit gate count. In particular, we focus on quantum algorithms essential for Shor's factoring algorithm \cite{Shor1997}, the quantum Fourier transform (QFT) \cite{NielsenChuang2011} and quantum addition algorithm based on the QFT \cite{Draper2000}, as well as the implementation of Grover's diffusion operator \cite{Grover1996}. Furthermore, we present a strategy to efficiently prepare graph states, which represent an important resource for measurement-based quantum computing \cite{Hein2006}.

\section{Common gates and gate sequences}\label{sec:common_gates}
\subsection{Arbitrary single-qubit gates}
\begin{figure}
    \centering
    \scalebox{0.6}{\includegraphics{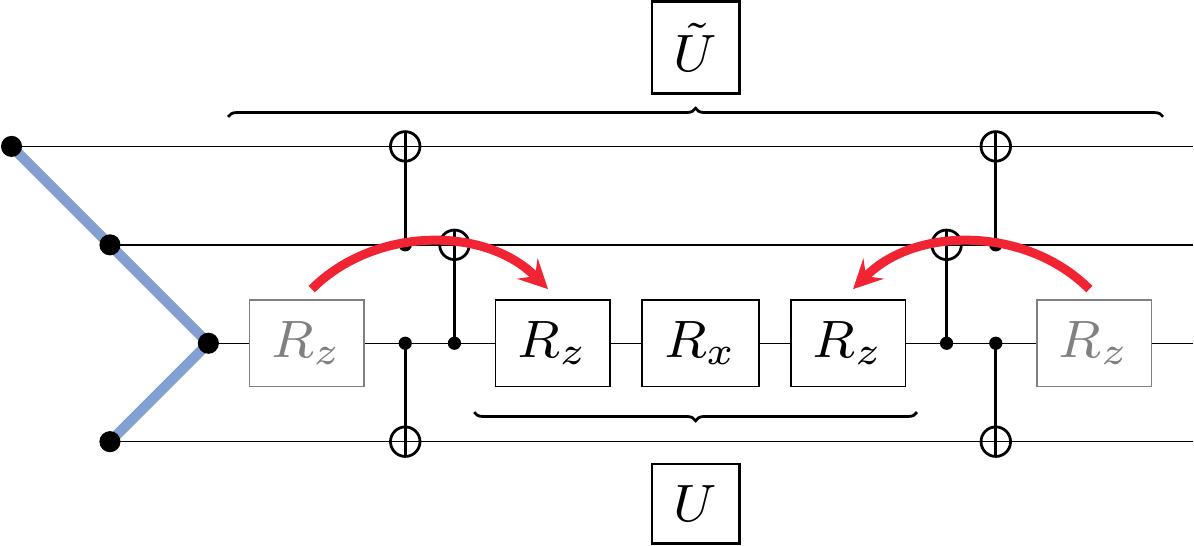}}
    \caption{Realization of a single-qubit unitary in universal parity quantum computation by exploiting Eq.~\eqref{eq:unitary_decomposition}. CNOT gates commute with $R_z$ gates acting on their control qubit. The $R_z$ gates on the data qubit can thus be moved through the CNOT sequence and recombined to the unitary $U$ together with the $R_x$ rotation. The unitary $U$ acts locally on a physical qubit, while $\tilde U$ denotes a logical operation. The blue line represents the logical line corresponding to the qubit targeted by the operation. }
    \label{fig:unitary_circuit}
\end{figure}

Any single-qubit unitary $U$ can be decomposed into rotations \cite{NielsenChuang2011}
\begin{equation}\label{eq:unitary_decomposition}
    U=R_z(\alpha)R_x(\beta)R_z(\gamma),
\end{equation}
with some angles $\alpha$, $\beta$, and $\gamma$.
We can thus construct any logical single-qubit gate using the operators defined in Eqs.~\eqref{eq:logical_rx} \eqref{eq:logical_rz} as
\begin{equation}
    \tilde{U}=\tilde R_z(\alpha)\tilde R_x(\beta)\tilde R_z(\gamma).
\end{equation}
The two $\tilde R_z$ rotations can be easily implemented in the LHZ scheme with physical rotations on the corresponding data qubits. The $\tilde R_x$ rotation requires a chain of controlled-NOT (CNOT) gates along the logical line and a physical $R_x$ rotation on one of its qubits, as discussed in Ref.~\cite{universality_paper}.
If the physical $R_x$ rotation is chosen to be performed on the respective data qubit, it can be recombined with the surrounding $R_z$ rotations to the unitary $U$, now acting on the physical qubit, as shown in Fig.~\ref{fig:unitary_circuit}. This is possible because the CNOT gates commute with $R_z$ gates acting on their control qubit.
Hence, a single-qubit unitary on a logical qubit $(i)$ can be implemented by performing the gate on the data qubit $(i)$ and applying CNOT gates along the corresponding logical line. 

Note that the single-qubit $R_x$ rotation can be performed on a parity qubit instead of the data qubit of the line. In that case, the decomposed rotations cannot be recombined, but for ${n>4}$, the additional $R_z$ rotations can be performed in parallel with the CNOT chains. In combination with a partial parallelization of CNOT chains (see the Appendix of Ref.~\cite{ender2022modular}), this leads to a minimal circuit depth for any single-qubit unitary of
\begin{equation}
    d_U = 2\left\lceil \frac{n}{2}\right\rceil + 1.
\end{equation}
This result can be seen from the implementation of an $\tilde R_x$ gate, as shown in Fig.~\ref{fig:cnot_depth} as a part of the logical CNOT gate. This implementation requires three single-qubit rotations (of which only one requires a separate step in depth) and ${2(n-1)}$ CNOT gates. 

For products of single-qubit gates, it can be beneficial to apply the decoding sequence, perform the rotations on the respective data qubit and encode again, which adds up to a circuit depth of ${2n+3}$, as discussed in Ref.~\cite{universality_paper}.

\subsection{Two-qubit gates}
\begin{figure}
    \centering
    \includegraphics[width=\columnwidth]{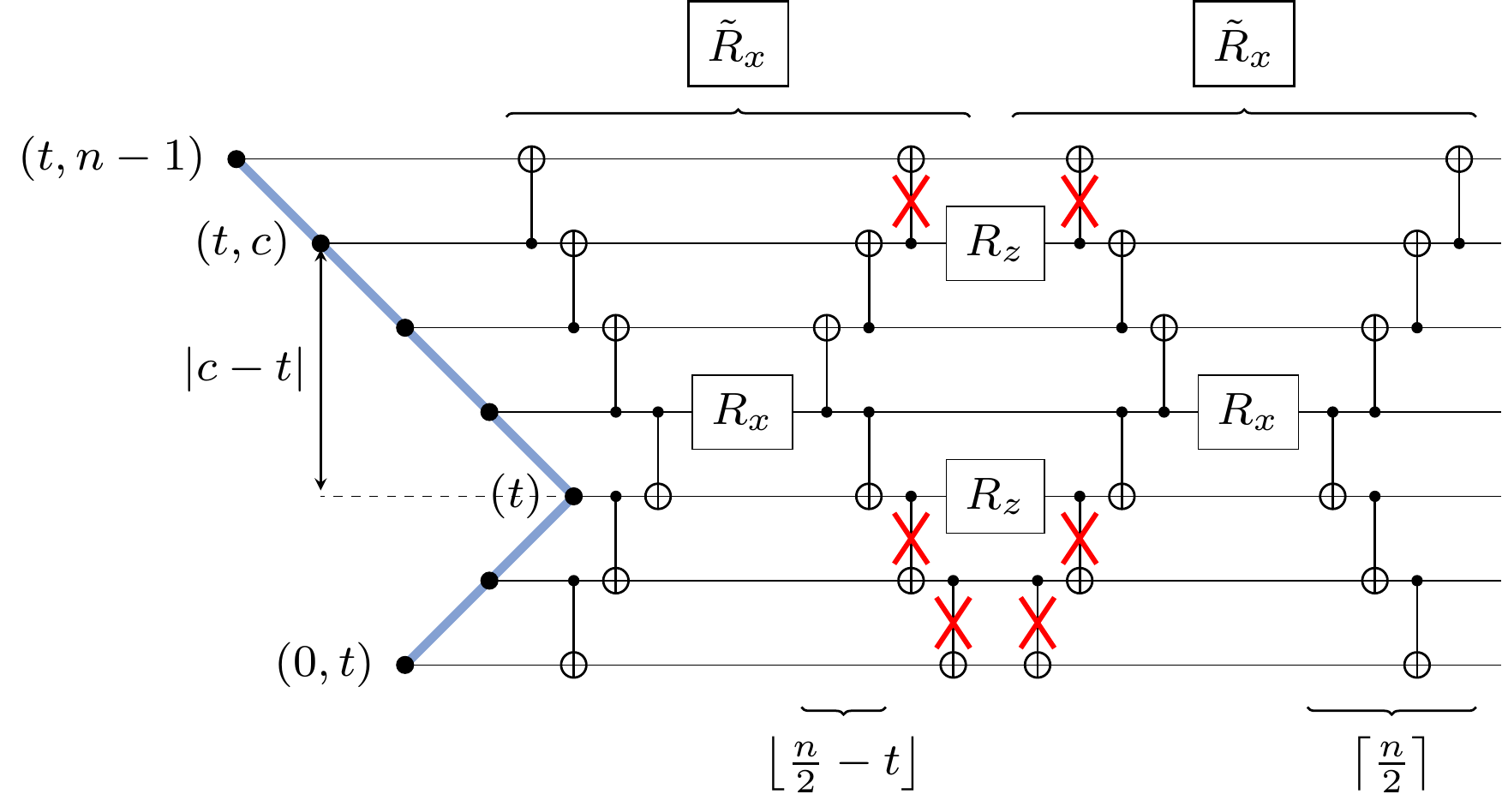}
    \caption{Implementation of a CNOT gate in the LHZ architecture. The gate count depends on the number of logical qubits $n$ and the distance ${|c-t|}$ between control ($c$) and target ($t$) qubits. CNOT gates marked with a red cross cancel; the blue line represents the logical line corresponding to the target qubit. The outer $R_z$ rotations of the Hadamard gates are not shown, the $R_z$ gates on qubit $(t)$ have been merged.}
    \label{fig:cnot_depth}
\end{figure}

While a logical $\text{C}\tilde{\text{P}}_\phi$ gate can be implemented in the parity encoding with only physical single-qubit gates according to Eq.~\eqref{eq:logical_cphase}, a logical CNOT gate requires two additional Hadamard gates on the target qubit,
\begin{equation}\label{eq:cnot_decomposition}
    \text{CNOT}^{(c, t)} = H^{(t)}\text{CZ}^{(c, t)}H^{(t)}.
\end{equation}
In the decomposition \eqref{eq:unitary_decomposition}, logical Hadamard gates require $\tilde{R}_x$ gates and thus CNOT chains along the logical line.
An implementation of the decomposition \eqref{eq:cnot_decomposition} in the LHZ scheme is depicted in Fig.~\ref{fig:cnot_depth} and results in a circuit depth of

\begin{equation}
    d_\text{CNOT}=2\left(\left\lceil\frac{n}{2}\right\rceil+\left\lfloor\max\left(\left|\frac{n}{2}-c\right|, \left|\frac{n}{2}-t\right|\right)\right\rfloor+k\right)+3,
\end{equation}
where $c$ and $t$ represent the index of the control  and the target qubit, respectively and
\begin{equation}
   k= \begin{cases}
    1 & \text{if}\, \left|\frac{n}{2}-c\right| = \left|\frac{n}{2}-t\right|\\
0 & \text{otherwise.}
    \end{cases}
\end{equation}
The circuit depth $d_\text{CNOT}$ corresponds to the depth of two $\tilde R_x$ gates minus the depth saved due to gate-cancelling, as depicted in Fig.~\ref{fig:cnot_depth}. The number of required (physical) CNOT gates is ${2(n-1+|c-t|)}$, which is less than the CNOT count for two logical Hadamard gates, also due to canceling gates. Furthermore, seven single-qubit rotations are needed to construct a logical CNOT gate (six for the Hadamard parts and three for the CZ part, where the $R_z$ rotations on the data qubit can be merged).

\begin{table}
    \centering
    \scriptsize
\begin{tabular}{c c c||c||c|c|c|c}

$s_i$ & $s_j$ & $s_k$ & $\text{CC}\tilde{\text{P}}_{\phi}^{(i,j,k)}$ & $\text{C}\tilde{\text{P}}_{\phi/2}^{(i, j)}$ & $\text{C}\tilde{\text{P}}_{\phi/2}^{(i,k)}$ & $\sigma_x^{(jk)}\text{CP}_{\phi/2}^{(i,(jk))} \sigma_x^{(jk)}$ & $P_{-\phi/2}^{(i)}$\tabularnewline
\hline 
\hline 
$0$ & $0$ & $0$ & $0$ & $0$ & $0$ & $0$ & $0$\tabularnewline
\hline 
$0$ & $0$ & $1$ & $0$ & $0$ & $0$ & $0$ & $0$\tabularnewline
\hline 
$0$ & $1$ & $0$ & $0$ & $0$ & $0$ & $0$ & $0$\tabularnewline
\hline 
$0$ & $1$ & $1$ & $0$ & $0$ & $0$ & $0$ & $0$\tabularnewline
\hline 
$1$ & $0$ & $0$ & $0$ & $0$ & $0$ & $\phi/2$ & $-\phi/2$\tabularnewline
\hline 
$1$ & $0$ & $1$ & $0$ & $0$ & $\phi/2$ & $0$ & $-\phi/2$\tabularnewline
\hline 
$1$ & $1$ & $0$ & $0$ & $\phi/2$ & $0$ & $0$ & $-\phi/2$\tabularnewline
\hline 
$1$ & $1$ & $1$ & $\phi$ & $\phi/2$ & $\phi/2$ & $\phi/2$ & $-\phi/2$\tabularnewline

\end{tabular}
    \caption{Phases acquired by application of the constituent operators for the $\text{CC}\tilde{\text{P}}_{\phi}^{(i,j,k)}$ gate on a computational basis state ${\ket{s_i}\ket{s_j}\ket{s_k}}$. For all basis states, the phases of the gate sequence sum up to the required phase. Physical operations, which are applied directly to the parity or data qubits, are written without a tilde, while a tilde denotes effective logical operations.}
    \label{tab:ccp_proof}
\end{table}

\subsection{Intrinsic higher-order interactions}
It is possible to encode the cumulative parity of multiple logical qubits $(q_i)$ in a single parity qubit, with operator correspondences
\begin{equation}
     \tilde\sigma_{z}^{(q_1)}\tilde\sigma_{z}^{(q_2)}\cdots\tilde\sigma_{z}^{(q_n)}\ket{\psi} = \sigma_z^{(q_1q_2\dots q_n)}\ket{\psi}
\end{equation}
for ${\ket{\psi}\in \mathcal{H}_\text{CF}}$. As an example, consider the three-body parity qubit $(ijk)$. With this qubit, a logical three-body interaction of the form
\begin{equation}
    \exp\left(i \phi \tilde\sigma_{z}^{(i)}\tilde\sigma_{z}^{(j)}\tilde\sigma_{z}^{(k)}\right)
\end{equation}
is directly accessible via the physical single-qubit operation
\begin{equation}
\exp\left(i \phi \sigma_z^{(ijk)}\right).
\end{equation}
Note that the placement of parity qubits $\sigma_z^{(q_1q_2\cdots q_n)}$ involved in more than two logical lines typically requires a tailored qubit layout \cite{ender2021compiler}. This approach is in particular useful for solving combinatorial optimization problems with higher-order interactions (see Sec.~\ref{sec:optimization_problems}).
\subsection{Derived higher-order interactions}
Alternatively, we can combine parity qubits with actual physical two-qubit gates to obtain logical multi-qubit gates.

Consider, for example, a physical controlled phase gate $\text{CP}_{\phi}^{(i, (jk))}$ between a parity qubit $(jk)$ and a data qubit $(i)$. For a computational basis state ${\ket{s_i}\ket{s_j}\ket{s_k}}$, ${s_{\{i, j, k\}}\in\{0, 1\}}$, this gate applies a phase if and only if ${s_j\neq s_k}$ and ${s_i=1}$. 
Flipping the parity qubit $(jk)$, this can be used to construct a logical 2-controlled phase gate (and in particular a CCZ gate for ${\phi=\pi}$) as
\begin{equation}\label{eq:ccp_decomposition}
\begin{split}
\text{C}&\text{C}\tilde{\text{P}}_\phi^{(i, j, k)} \\&=\text{C}\tilde{\text{P}}_{\phi/2}^{(i, j)}\, \text{C}\tilde{\text{P}}_{\phi/2}^{(i, k)}\, \sigma_x^{(jk)}\,\text{CP}_{\phi/2}^{(i, (jk))}\, \sigma_x^{(jk)}\, P_{-\phi/2}^{(i)}.
\end{split}
\end{equation}
Here, $P_{-\phi/2}^{(i)}$ is a single-qubit phase gate and up to a global phase equivalent to a single-qubit rotation $R_z^{(i)}(-\phi/2)$.
The action of the operators in Eq.~\eqref{eq:ccp_decomposition} on all computational basis states is given in Table~\ref{tab:ccp_proof}.\\

The Toffoli gate, an important constituent of many quantum circuits \cite{Shor1997, Vedral1996Arithmetic, Figgatt_2017_3qubitGrover, Dennis2001FaultTolerant}, can be decomposed as
\begin{equation}\label{eq:ccnot_decopmosition}
    \text{CCNOT}^{(i, j, k)} = H^{(k)} \text{CC}Z^{(i, j, k)} H^{(k)}
\end{equation}
and, together with Eq.~\eqref{eq:ccp_decomposition}, reduced to two-qubit operators in the parity scheme. Here, $i$ and $j$ denote the control qubits and $k$ the target qubit. The circuit depth and the two-qubit gate count of this decomposition depend solely on the number of parity qubits in the logical line of the target qubit (i.e.\ the connectivity requirements of the logical qubit $k$). All required gates apart from the Hadamard gates on the target qubits can be implemented natively (i.e.\ without CNOT chains) as long as the data qubit $(i)$ and the parity qubit $(jk)$ are close enough on the chip to perform the physical controlled phase gate.

Similarly, a physical $\text{CP}_\phi$ gate between two parity qubits $(ij)$ and $(kl)$ effectively represents a four-qubit gate which adds a phase if and only if ${s_i\neq s_j}$ and ${s_k\neq s_l}$. 
In the square lattice implementation of the all-to-all connected graph, these derived multi-qubit gates can be realized along the diagonal of a parity constraint.
If the required connectivity is not available, the qubits can be rearranged to an algorithm-specific layout to provide the desired interactions.

\subsection{Negative controls}
For many applications \cite{Gado2021OptimizationNegativeControl, Rahman2014TemplatesToffoli}, it is useful to invert control inputs of controlled multi-qubit gates. That corresponds to applying a spin-flip on the respective qubit before and after the controlled gate. In the parity encoding, a logical spin flip is realized by simultaneously flipping all physical spins along a logical line \cite{universality_paper}. However, as the spin flips on parity qubits not involved in the operation cancel out, it is sufficient to only flip the data qubit corresponding to the desired negative control and the parity qubits involved in decomposition \eqref{eq:ccp_decomposition}.

\begin{figure*}
    \centering
    \includegraphics[width=\textwidth]{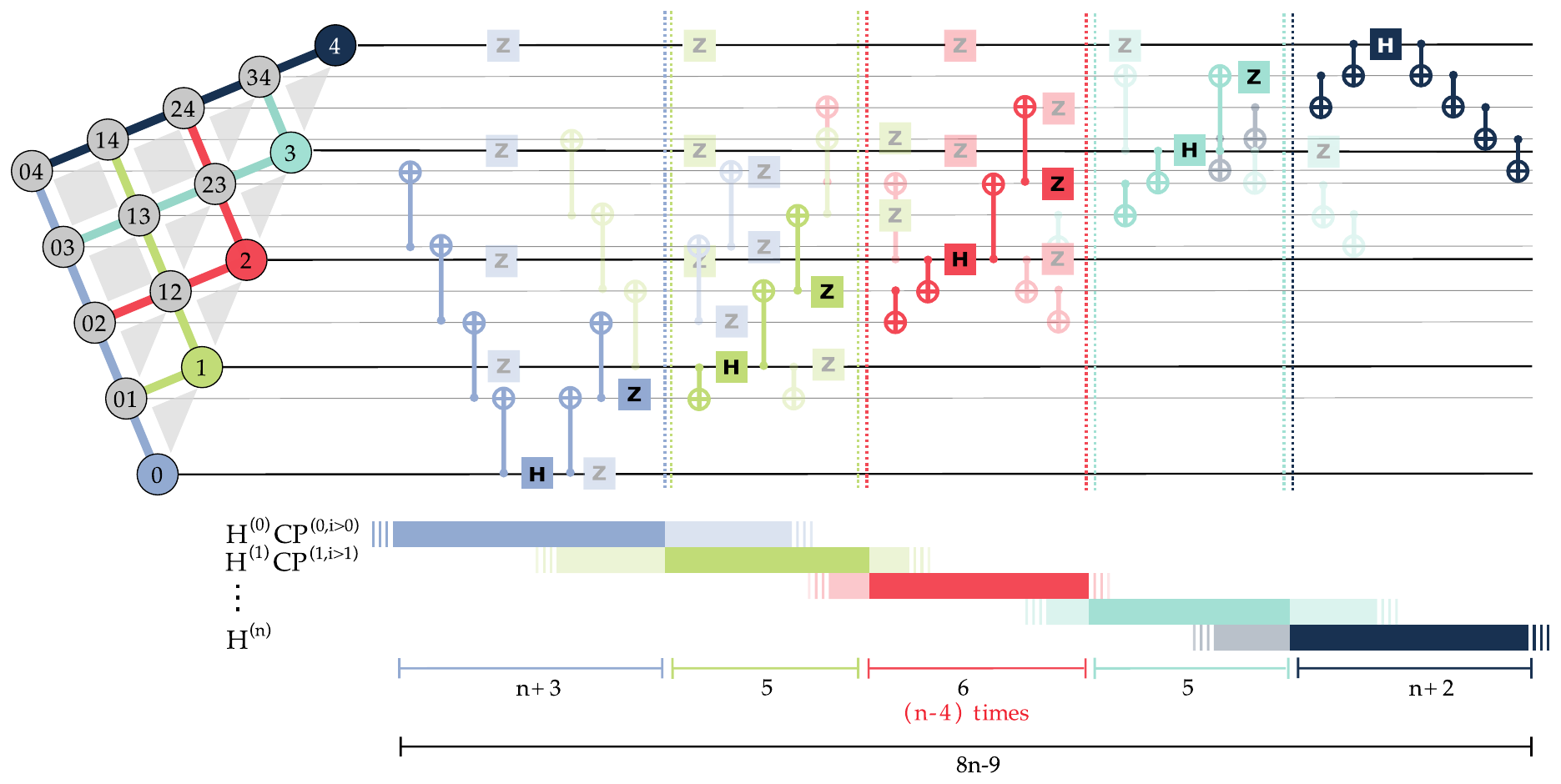}
    \caption{Circuit for performing the QFT in the LHZ scheme. CNOT gates are required to perform the logical Hadamard gates, while single-qubit $R_z$ rotations implement logical CPhase gates. Gates which do not contribute to the circuit depth are faded out. For all but the first and the last line, increasing the system size would only add gates which do not contribute to the circuit depth. The diagram on the bottom visualizes the time steps occupied by each gate block of the QFT algorithm. In the layout on the left, lines depict logical lines and the gray squares and triangles represent parity constraints.}
    \label{fig:qft}
\end{figure*}

\section{Applications}
\subsection{Quantum Fourier transform}\label{sec:qft}
The quantum Fourier transform \cite{NielsenChuang2011} is the quantum analog to the discrete Fourier transform and an important ingredient for many quantum algorithms such as quantum phase estimation \cite{Kitaev1996} and, in further consequence, Shor's factoring algorithm \cite{Shor1997}. The unitary for the QFT on $n$ qubits is given by
\begin{equation}
    U_\text{QFT} = \prod_{i=1}^n \left[ H^{(i)}\prod_{j=i+1}^n\text{C}R^{(i, j)}_{j-i+1}\right],
\end{equation}
where the gates $R_k$ are in this context
defined as
\begin{equation}
    R_k \equiv \text{P}_{2\pi/2^k} = 
    \begin{pmatrix}
        1 & 0 \\
        0 & e^{2\pi i /2^k}
    \end{pmatrix}
\end{equation}
and the superscripts denote the logical qubits the unitary is acting on. For an implementation of the QFT on a square lattice device in the standard gate model, SWAP gates are required to realize the controlled-phase gates. In contrast to that, in the parity architecture, these can be completely removed at the cost of requiring CNOT gates to perform the Hadamard gates according to the decomposition \eqref{eq:unitary_decomposition}. The corresponding circuit for a possible low-depth implementation is shown in Fig.~\ref{fig:qft}. The CNOT chains for implementing the logical Hadamard gates can be parallelized apart from a small contribution: For the logical qubit ${(0)}$ (light blue line in the figure), the first CNOT chain along the line, the Hadamard gate on the data qubit and the second chain until qubit ${(0,1)}$ need to be performed before qubit ${(0, 1)}$ is free to be used in the next line. In general, for any line $i$, only the gates between the crossing to the previous line [qubit ${(i-1, i)}$] and the crossing to the next line [qubit ${(i, i+1)}$] contribute to the circuit depth. The necessary logical CPhase gates appear in the circuit as single-qubit $R_z$ rotations (marked with $Z$). Note that many of them can be merged into other $R_z$ rotations and are therefore not shown in Fig.~\ref{fig:qft}. Following that procedure, the operations on the first qubit (light blue) block ${n+3}$ steps. The second and the ${(n-1)}$th ones take five steps, and the last one (dark blue) ${n+2}$, because there is no CPhase gate required. All other operations in between occupy six time steps. This construction results in a total circuit depth of ${8n-9}$.

A comparison of the required resources in the parity architecture and in the standard gate model is given in Table~\ref{tab:qft_gatecount}. The circuit depth for the QFT in the LHZ scheme is up to a constant the same as in an all-to-all connected setup. This is remarkable in that the LHZ scheme can be implemented in current experiments and does not rely on impracticable hardware requirements.
\begin{table}
    \centering
    \begin{tabular}{c|c|c|c}
         &  all-to-all & square lattice & LHZ\\ \hline
    qubits & $n$ & $n$ &$\frac{1}{2}n(n+1)$\\
    CNOT & $n(n-1)$ & $\frac{3}{2}n(n-1)$ & $2n(n-1)$ \\
    single-qubit & $n^2$ & $n^2$ & $\frac{1}{2}n(n+3)$\\
    total gates & $n(2n-1)$ & $\frac{1}{2}n(5n-3)$ & $\frac{1}{2}n(5n-1)$\\
    circuit depth & $8n-10$ & $10n-13$ & $8n-9$
    \end{tabular}
    \caption{Comparison of the required resources for the QFT implemented on an all-to-all connected device, a square lattice with nearest-neighbor interactions (requiring SWAP operations), and parity mapped on a square lattice with nearest-neighbor interactions. The numbers for the gate model implementations are taken from Ref.~\cite{Holmes2020_impact}, while the numbers for the LHZ encoding are analytically deduced.
    All operations have been decomposed into single-body rotations $R_x$, $R_z$, $H$, and CNOT gates.}
    \label{tab:qft_gatecount}
\end{table}

\subsection{Quantum Addition}
\begin{figure}
    \centering
    \scalebox{0.82}{\includegraphics{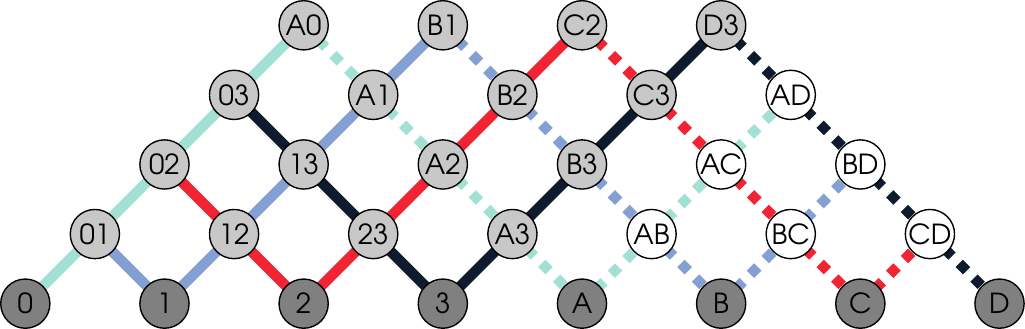}}
    \caption{Implementation proposal for the quantum addition of two registers R1 (labeled with numbers) and R2 (labelled with letters) in the LHZ scheme. The embedding consists of three standard LHZ schemes, two encoding the interactions within register R1 and R2, respectively, and one encoding the interregister interactions. The solid lines represent logical lines for register R1 and the dashed lines the logical lines for qubits in register R2. The qubits encoding interactions within register R2 (white) can be left out if no operations are necessary in register R2.}
    \label{fig:quantum_addition_lhz}
\end{figure}
Another prominent problem in various quantum algorithms is
quantum addition as the basic building block of algebraic manipulation of quantum registers.
An efficient circuit for the addition of two quantum registers based on the QFT, proposed by Draper \cite{Draper2000}, performs the addition in the Fourier space with conditional rotation gates.
Reference~\cite{Ruiz-Perez2017} suggests extensions of the resulting circuit that allow for computing the mean or weighted sum of a set of numbers, or even performing (modular) multiplication and exponentiation. 

Quantum addition (modulo $2^n$) of two $n$-qubit quantum states stored in registers R1 and R2 is realized by performing a QFT on one register (say, register R1), then performing controlled $R_z$ rotations between the two registers and finally applying the inverse QFT to the previously Fourier transformed register \cite{Draper2000, Ruiz-Perez2017}. The corresponding quantum circuit is given by
\begin{equation}
    U_\text{QFT}^\text{R1} \left[\prod_{i=0}^{n-1}\prod_{j=i}^{n-1} \text{C}R_{j-i+1}^{(b_{n-j}, a_{n-i})}\right] U_\text{QFT}^{\text{R1}\dagger},
\end{equation}
where $a_i$ and $b_i$ denote qubits of registers R1 and R2, respectively, and $U_\text{QFT}^\text{R1}$ denotes a QFT on register R1. An illustration of the circuit is provided in Ref.~\cite{Draper2000}.

On a chip with all-to-all connectivity, the quantum addition algorithm can be implemented in logarithmic depth, neglecting the gates required for performing the necessary QFT-steps. In the parity encoding, we can perform the logarithmic-depth part in a single time step and add the linear-depth QFT circuits, without requiring any SWAP gates.

As discussed in Sec.~\ref{sec:common_gates}, the controlled rotations can be implemented with single-qubit operations only, provided the necessary parity qubits are available. In order to fulfill that requirement, we suggest the qubit layout depicted in Fig.~\ref{fig:quantum_addition_lhz}. The amount of physical qubits compared to the previously introduced architecture increases by ${n(n+1)}$, such that there are ${3n(n+1)/2}$ qubits in total.

If one of the registers does not require any computations within itself (for example if it represents classical data), the parity qubits corresponding to interactions within that register are not necessary, and the data qubits of the register can be added directly in continuation of the remaining logical lines, requiring $n(n+2)$ qubits in total.
Note that either way, the extensions of logical lines hardly affect the circuit depth of the QFT part, as they do not cross any lines corresponding to the other register. Our architecture therefore allows us to perform the core quantum addition circuit in a single time step, and the surrounding QFT parts with a depth linear in $n$, as discussed in Sec.~\ref{sec:qft}. A list comparing required resources for the quantum circuit in the standard gate model and in the parity scheme is given in Table~\ref{tab:quantum_addition_gatecount}.

\begin{table}
\centering
\begin{tabular}{c|c|c}
Resource      & all-to-all        & LHZ\\ \hline
qubits        & $2n$                 & ${n(n+2)}$    \footnote{If computations involving multi-qubit gates are required within both registers, the number of qubits increases to $\frac{3}{2}n(n+1)$.} \\
CNOT          & $n(n+1)$             & $0$\\
single-qubit  & $\frac{1}{2}n(n+5)$  & $n(n+2)$\\
total gates   & $\frac{1}{2}n(3n+7)$ & $n(n+2)$\\
circuit depth & $3\log_2(n)+1$       & $1$
    \end{tabular}
    \caption{Required resources for the core step of the QFT-based Quantum Addition algorithm on an all-to-all connected device, and parity mapped on a square lattice with nearest-neighbor connectivity. The resources required for the involved QFT-circuits are not included in this table. All operations have been decomposed into single-body rotations $R_x$ and $R_z$, $H$, and CNOT gates.}
    \label{tab:quantum_addition_gatecount}
\end{table}

\subsection{Multi-controlled Gates and Grover's Diffusion operator}
\begin{figure}
    \centering
    \includegraphics{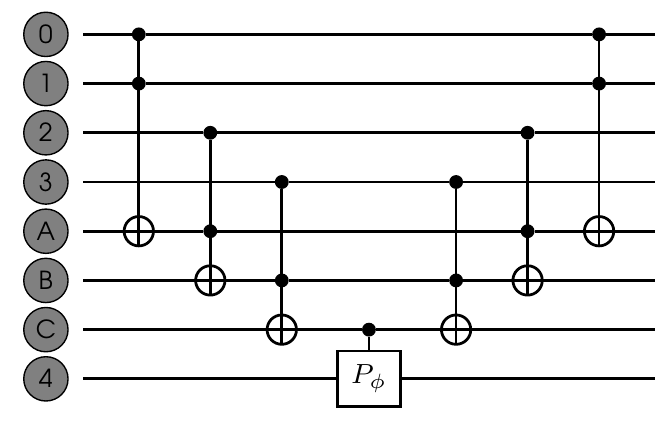}
    \caption{Gate decomposition of a multi-controlled phase gate into Toffoli and $\text{CP}_\phi$ gates using ancilla qubits $A$--$C$, following Ref.~\cite{NielsenChuang2011}.}
    \label{fig:n-toffoli}
\end{figure}

\begin{figure}
    \centering
    \includegraphics{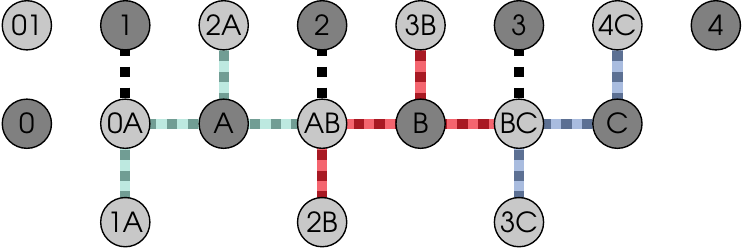}
    \caption{Implementation layout for the n-controlled phase gate. Ancilla qubits are labeled with capital letters, parity qubits are shown in light gray, and data qubits are dark gray. Dotted lines indicate connectivity requirements between the physical qubits during computation. The parity encoding is reduced to the minimal required set of physical qubits, such that only the ancilla qubits require (short) logical lines, shown by the colored lines.}
    \label{fig:grover_implementation}
\end{figure}
A difficulty arising in many quantum algorithms is the implementation of multi-controlled quantum gates, as it requires a considerable number of nonlocal interactions and therefore SWAP gates. In the following, we present an implementation of the $m$-controlled phase gate on a parity-mapped architecture with ${m+1}$ logical qubits. In Grover's search algorithm \cite{Grover1996}, the multicontrolled phase gate is especially relevant for the implementation of the diffusion operator \cite{Zhang_2020_diffusion_depth}. The diffusion operator on ${m+1}$ qubits corresponds to an $m$-controlled phase gate. We exploit the decomposition into 2-controlled phase and Toffoli gates \cite{NielsenChuang2011}, which are further decomposed according to Eq.~\eqref{eq:ccnot_decopmosition}. The decomposition is depicted in Fig.~\ref{fig:n-toffoli} and introduces ${m-1}$ ancilla qubits, labeled with capital letters. Here, $m$ is the number of control qubits. In the LHZ scheme, this enables us to implement an $m$-controlled phase gate $\text{C}^m\tilde{\text{P}}_\phi$ (and multicontrolled-NOT gates by using Hadamard gates) with gate resources scaling linearly with $m$ and introducing ${4m+3}$ ancilla qubits in total. 
As discussed in Ref.~\cite{universality_paper}, it is not necessary to use all parity qubits if the corresponding connections are not required. Figure~\ref{fig:grover_implementation} shows a possible qubit arrangement to implement the diffusion operator with the introduced decomposition for ${m=4}$. In this layout, the $\text{CC}\tilde{\text{P}}_\phi^{(i, j, k)}$ gates are decomposed such that the physical $\text{CP}_{\phi/2}^{(i, (jk))}$ gate occurring in Eq.~\eqref{eq:ccp_decomposition} is performed between a problem qubit $(i)$ and an ancilla qubit $(jk)$. On the chip layout in Fig.~\ref{fig:grover_implementation}, these correspond to the black dashed lines. The other $\text{CP}_\phi$ gates are performed via the protocol given in Eq.~\eqref{eq:logical_cphase}. 

The Hadamard gates in Eq.~\eqref{eq:ccnot_decopmosition}, which only occur on ancilla qubits, can be implemented with eight CNOT gates each, as the logical lines (colored lines in Fig.~\ref{fig:grover_implementation}) of ancilla qubits consist of only five qubits, independent of the number of control qubits. This allows for an implementation of the $m$-controlled-NOT gate scaling linearly in circuit depth as well as in gate count. We note that this example serves as a proof of principle for the implementation of Grover's diffusion operator in the parity encoding. While it cannot beat standard implementations using, e.g., Margolus construction of a Toffoli gate \cite{Barenco1995_Elementary} in terms of quantum resources, the encoding offers additional flexibility, especially regarding chip layouts with sparse connectivity.

\subsection{Optimization Problems}\label{sec:optimization_problems}
Optimization problems can be formulated as energy minimization of Hamiltonians of the form
\begin{equation}\label{eq:problem_hamiltonian}
    H_\text{Z} = \sum_{i} J_{i} \sigma_z^{(i)} +  \sum_{i,j} J_{ij} \sigma_z^{(i)} \sigma_z^{(j)} 
    + \sum_{i,j,k} J_{ijk} \sigma_z^{(i)} \sigma_z^{(j)} \sigma_z^{(k)} + \cdots
\end{equation}
with the particular optimization problem being encoded in the pre-factors $J_i$, $J_{ij}$, etc.
Such optimization problems can be tackled on digital, gate-based architectures using the quantum approximate optimization algorithm (QAOA) \cite{Farhi2014}. The QAOA variationally evolves the system with Hamiltonians $H_\text{X}$ and $H_\text{Z}$ as
\begin{equation}\label{eq:qaoa_state_lhz}
\ket{\psi} = \prod_{j=1}^p e^{-i\beta_j {H}_\text{X}}e^{-i\gamma_j {H}_\text{Z}}
\end{equation}
where $H_\text{X}=\sum_i \sigma_x^{(i)}$ is a driver term to explore the search space. While the problem Hamiltonian $H_\text{Z}$ can be implemented in the parity architecture in a single step, the driver Hamiltonian can be realized with a gate sequence of depth ${7n-8}$ analogous to the circuit shown in Fig.~\ref{fig:qft} (leaving out the $R_z$ rotations and replacing the physical Hadamard gates by parameterized $R_x$ rotations). 
Such implementations and similar variants have been analyzed thoroughly in recent literature \cite{lechner2020, ender2022modular, Rocchetto2016_stabilizer}. 
Typically, an additional Hamiltonian containing parity constraints is added to give an energy penalty to states which are not in the code space. These constraints simplify the operations required for the driver Hamiltonian. In particular, whenever all parity constraints are energetically penalized, the driver Hamiltonian reduces to a sum of single-body terms.

\subsection{Preparation of Graph States}
Graph states are an important resource for measurement-based quantum computing \cite{Raussendorf2001, Raussendorf2003} as well as for several error correction protocols \cite{Raussendorf2007, Schlingemann2001}.
Cluster states, corresponding to square grid graphs, can be prepared efficiently using nearest-neighbor interactions. The preparation of more arbitrary graph states, however, typically requires long-range interactions between many different qubits, as it involves applying a CZ gate for every edge in the graph \cite{Hein2006}. In the parity encoding, arbitrary graph states containing $n$ qubits can be created with a circuit depth of at most ${n+3}$ and a CNOT-gate count of ${2n(n-1)}$ as follows. 
We start by preparing $n$ data qubits in the superposition state $\ket{+}$. We then apply the encoding sequence, in either the LHZ scheme or a reduced version of it, depending on the connectivity requirements of the respective graph, requiring a circuit of depth ${n+1}$ or less (for reduced versions). Subsequently, we apply CZ gates to every pair of logical qubits to be connected in the graph, using parallel single-qubit gates on the physical qubits according to the decomposition~\eqref{eq:logical_cphase}. 
The result of this is a graph state encoded in the parity architecture.
For measurement based quantum computing, which involves measurements of the logical qubits (in arbitrary axes), an additional decoding step of depth $n+1$ is required.

The number of parity qubits required for this procedure is in principle equal to the number of edges in the desired graph state. However, in some cases it can be useful to have additional ancillary qubits to simplify the encoding and decoding circuits.

\section{Conclusion and Outlook}
In this work, we have presented an approach to implement fundamental quantum algorithms for arbitrary system sizes, completely avoiding SWAP gates. A gate count analysis shows that our implementation has the potential to save multiqubit gates or reduce circuit depth for key constituents of Shor's algorithm.
In addition, tailored parity encodings for particular algorithms can be constructed by utilizing the parity compiler \cite{ender2021compiler}.
On top of that, due to the redundant encoding of information, the parity encoding provides an intrinsic potential to detect and correct bit-flip errors.

One possible approach to error correction in the parity encoding is the belief propagation proposed in Ref.~\cite{Pastawski2016}. The effects of this fault tolerance on circuit execution results was not covered in this work.

Our proposal requires interactions between nearest neighbors only, and can thus be implemented on various current NISQ devices with their natural inter-qubit connectivity. The proposal is also  independent of the specific qubit platform.
Suitable platforms are for example superconducting qubits~\cite{wallraff2004strong, Koch2007, Houck2008, Barends2013, arute2019quantum, Wu2021}, neutral atoms \cite{saffman2010quantum, henriet2020quantum, rydberg_cong_2022}, or trapped ions~\cite{blatt2012quantum, kielpinski2002architecture, Lekitsch2017, ion_hughes_2020}.
In order to complement the bit-flip tolerance of the parity encoding, the use of noise-biased qubits \cite{Aliferis2009, Puri2020, Lescanne2020} may be considered.

A combination of our findings regarding the QFT in the LHZ scheme with the achievements presented in Refs.~\cite{Sieberer2018, Dlaska_2019} on programming arbitrary superposition states using quantum annealers may give rise to a QFT device without exponential gate overhead for the initial state preparation \cite{NielsenChuang2011} and opens up a promising avenue for further research.\\

\paragraph*{Acknowledgements -} We thank C. Ertler for valuable input on the graph state preparation and for numerous fruitful discussions. Work at the University of Innsbruck is supported by the Austrian Science Fund (FWF) through a START grant under Project No. Y1067-N27 and the Special Research Programme (SFB) ``BeyondC: Quantum Information Systems Beyond Classical Capabilities'', Project No. F7108-N38. This work was supported by the Austrian Research Promotion Agency under Grant (FFG Project No. 892576, Basisprogramm).

%

\end{document}